\documentclass{astro-ph-1}
\usepackage{graphicx}
\usepackage{mathptmx}
\newcommand\ltsim{\,\lower0.7ex\hbox{$\stackrel{<}{\sim}$}\,}
\newcommand\gtsim{\,\lower0.7ex\hbox{$\stackrel{>}{\sim}$}\,}

\begin{document}
\title{Multigroup Models of the Convective Epoch \\ in Core Collapse 
Supernovae
\footnote{\uppercase{i}nvited talk, presented at \uppercase{S}ci\uppercase{DAC} 2005, \uppercase{S}an \uppercase{F}rancisco, \uppercase{CA, USA},
26--30 \uppercase{J}une 2005; to appear in {\em \uppercase{J}ournal of \uppercase{P}hysics: \uppercase{C}onference \uppercase{S}eries.}}
}

\author{F. Douglas Swesty and Eric S. Myra}

%\address{Dept. of Physics and Astronomy \\
%State University of New York at Stony Brook \\
%Stony Brook, NY \mbox{ 11794--3800}, USA \\
%
%\ead{
%dswesty@mail.astro.sunysb.edu, 
%emyra@mail.astro.sunysb.edu
%}
\address{Dept. of Physics \& Astronomy \\
State University of New York at Stony Brook \\ 
Stony Brook, NY \mbox{ 11794--3800} \mbox{ USA}\\ 
\begin{tabular}{ll}
E-mail:  & dswesty@mail.astro.sunysb.edu\\
&  emyra@mail.astro.sunysb.edu
\end{tabular}}

\maketitle
\fontfamily{ptm}

%\begin{abstract}
\abstracts{
Understanding the explosion mechanism of core collapse 
supernovae is a problem that has plagued nuclear astrophysicists
since the first computational models of this phenomenon
were carried out in the 1960s.  Our current theories of this 
violent phenomenon center around multi-dimensional effects 
involving radiation-hydrodynamic flows of hot, dense matter and
neutrinos.  Modeling these multi-dimensional radiative flows
presents a computational challenge that will 
continue to stress high-performance computing beyond
the teraflop to the petaflop level.  In this paper we describe
a few of the scientific discoveries that we have made via terascale
computational simulations of supernovae under the 
auspices of the SciDAC-funded Terascale Supernova Initiative.
}
%\end{abstract}.

\section{Introduction: Scientific and Computational Challenges}

For over four decades, researchers have struggled with one of the
great unsolved problems of astrophysics: how massive stars end their
lives via collapse of their cores and generate the explosions
we observe as supernovae.

The problem is an important one for diverse branches of physics.  The
supernova process is ultimately responsible for the distribution of
nearly all elements in the universe heavier than carbon.  Supernovae
frequently leave a compact remnant whose densities are at or above
nuclear densities. The majority of energy released during
gravitational collapse is imparted to weakly interacting
neutrinos. Finally, a collapse departing from spherical symmetry can
generate gravitational waves. Therefore, supernovae serve as natural
laboratories for matter at the highest densities and sit at the
frontier of research in astronomy and nuclear, particle, and
gravitational physics.

The supernova problem can be briefly stated.  Through nuclear fusion,
stars more massive than about 10~solar masses ($M_\odot$) evolve to
where the center consists of an iron core surrounded by layers of
successively lighter elements.  There is no exothermic reaction that
can process iron, and the core becomes dynamically unstable as a
result of electron capture by protons.  Collapse ensues, but
decelerates once nuclear density is exceeded because of the strong
repulsive force of nucleons.  The core bounces and rebounds, producing
a shock wave that starts propagating outwards through the mantle.  Were
the shock to continue moving in this fashion, matter would be
explosively ejected, and the supernova problem would be solved.
Instead, however, the most realistic supernova models collapse and
rebound, but create a shock wave that doesn't eject matter, either on
the hydrodynamic timescale ($\sim$~10~ms) or the diffusive timescale
of the escaping neutrino radiation ($\sim$~1--10~s).

This longstanding supernova puzzle has a longstanding history of
challenging the highest-performance computer systems of each technology
generation.  The difficulty of assembling a convincing supernova model
has led researchers to add increasingly sophisticated and
computationally intensive physics to the mix. Yet, despite great
advances in the field, supernova models continue to be oversimplified
and leave many relevant branches of physics inadequately explored.
The SciDAC program has enabled physicists to make substantial progress
in closing this gap and has permitted both fundamental research as
well as software development, allowing the field to progress at a more
optimistic pace.

In this article, we present recent results of core-collapse
simulations performed as part of the Terascale Supernova Initiative
under SciDAC.  These simulations were performed using 1024 IBM POWER-3
processors on {\em seaborg}, the IBM SP at the National Energy
Research Scientific Computing Center (NERSC).  Typical simulations
consumed between 50,000 and 100,000 processor-hours to advance 30~ms
of model time.

These are the first supernova simulations performed in a
two-dimensional geometry that also include a detailed multigroup
treatment of neutrino transport.  Our current focus is the 30 ms
immediately following core bounce.  This epoch gives rise to dynamic
instabilities in areas behind the stalled shock and leads to complex
convective behavior of the stellar fluid.  This is also a time when
neutrino radiation plays a major role in determining the all-important
transport of energy within the collapsed core.

\section{Our Model for a Core-Collapse Supernova Simulation}

Our model includes a radiation-hydrodynamic simulation algorithm to
solve the equations of (1)~hydrodynamic and (2)~neutrino-radiation
evolution.  The model also requires input of (3)~initial conditions,
including a pre-collapse iron stellar core, together with several of
the surrounding layers of lighter elements; (4)~an equation of state
to describe the behavior of nuclear matter over the diverse density,
temperature, and composition regimes that are encountered; and
(5)~neutrino microphysics to describe the weak interactions of
expected importance.

The computer code that implements this model, V2D, is a
two-dimensional, pure Eulerian, staggered-mesh code, loosely based on
the ZEUS-2D algorithm by Stone and Norman \cite{sn1,sn2,sn3}, but
considerably extended to permit its more general use in radiation
hydrodynamic problems.

V2D, an entirely new implementation coded according to the Fortran~95
standard, includes the design goals of portability, scalability,
componentization, and adherence to standards. Designed for
distributed-memory parallel architectures, V2D uses MPI-1 for message
passing between application tasks.  The SciDAC program has largely
enabled the development of V2D's high degree of scalability.  The
range of system sizes on which it currently runs includes standalone
Linux-based laptops and 2048 processors of {\em seaborg}, the IBM
POWER-3-based SP currently installed at NERSC. V2D formats data input
and output using parallel HDF5, built on the MPI-I/O portion of the
MPI-2 standard.

Consistent with our goal of componentization, we insist on a complete
separation of microphysics from the numerical implementation of our
radiation-hydrodynamics algorithm. This isolation of physics
components from the mathematics and computational-science components
has resulted in significant algorithmic and code improvements
contributed by our non-physicist colleagues.

\subsection{Hydrodynamics}

V2D's hydrodynamic component is solved in Newtonian formalism, wherein
the Eulerian equations are explicitly differenced.  For core collapse,
the problem is posed in spherical polar geometry and expressed in two
spatial dimensions by assuming azimuthal symmetry.  To accomplish
this, we use the following set of hydrodynamic equations:

\begin{equation} \label{eq:cont}
\frac{\partial \rho}{\partial t} +
{\pmb \nabla} \cdot \left( \rho {\bf v} \right) = 0
\end{equation}
\begin{equation} \label{eq:ye}
\frac{ \partial \left( \rho Y_e \right) }{\partial t} +
{\pmb \nabla} \cdot \left( \rho Y_e {\bf v} \right) = 
-m_b \sum_f \int d\epsilon 
\left( \frac{{\mathbb S}_{\epsilon}}{\epsilon} -
\frac{ \bar{{\mathbb S}}_{\epsilon}}{\epsilon} \right)
\end{equation}
\begin{equation} \label{eq:energy}
\frac{ \partial E}{\partial t} +
{\pmb \nabla} \cdot \left( E {\bf v} \right) +
P {\pmb \nabla} \cdot {\bf v} =
- \sum_f \int d\epsilon 
\left( {\mathbb S}_{\epsilon} + \bar{{\mathbb S}}_{\epsilon} \right)
\end{equation}
\begin{equation} \label{eq:mom}
\frac{ \partial \left( \rho {\bf v} \right) }{\partial t} +
{\pmb \nabla} \cdot \left( \rho {\bf v}{\bf v} \right) +
{\pmb \nabla} P + \rho {\pmb \nabla} \Phi +
{\pmb \nabla} \cdot 
\left\{ \sum_f \int d\epsilon \left( {\mathsf P}_{\epsilon}
+ \bar{{\mathsf P}}_{\epsilon} \right) \right\} 
= 0.
\end{equation}
Equation (\ref{eq:cont}) is the continuity equation for mass, where
$\rho$ is the mass density and ${\bf v}$ is the matter velocity, and
where these quantities, and those in the following equations, are
understood to be functions of position {\bf x} and time $t$. Equation
(\ref{eq:ye}) expresses the evolution of electric charge, where $Y_e$
is the ratio of the net number of electrons over positrons to the
total number of baryons.  In the presence of weak interactions, the
right-hand side is non-zero to account for reactions where the number
of electrons can change.  Here, we express the net emissivity of a
neutrino flavor (of energy $\epsilon$) and its antineutrino by
${\mathbb S}_{\epsilon}$ and $\bar{{\mathbb S}}_{\epsilon}$,
respectively.  This expression is integrated over all neutrino
energies and summed over all neutrino flavors $f$.  The mean baryonic
mass is given by $m_b$.  Evolution of the internal energy of the
matter is given by the gas-energy equation, Eq.~(\ref{eq:energy}),
where $E$ is the matter internal energy density and $P$ is the matter
pressure.  Again, the right-hand side of this equation is non-zero
whenever energy is transferred between matter and neutrino radiation
as a result of weak interactions.  Finally, Eq.\ (\ref{eq:mom})
expresses gas-momentum conservation, where $\Phi$ is the gravitational
potential, and ${\mathsf P}_{\epsilon}$ and $\bar{{\mathsf
P}}_{\epsilon}$ are radiation-pressure tensors for each energy and
flavor of neutrino and its anti-neutrino, respectively.

The combined use of explicit hydrodynamics, spherical polar
coordinates, and a spatial domain that includes the origin presents a
numerical stability problem. This follows from the degeneracy of the
coordinate system at $r=0$.  Clearly, all central zones at the vertex
should be in mutual instantaneous sonic contact.  However, since
standard explicit nearest-neighbor finite-difference techniques do not
include the coupling of all zones at the vertex, we introduce
numerical ``baffles'' into the center of the collapsed core, as though
it were a tank of fluid.  These baffles prevent both the angular
motion of the fluid inside the baffle radius and the severe
timestep-restricting Courant-Friedrichs-Levy (CFL) condition in the
$\theta$ direction near the core's center as a result of the
converging zones.  Zones on either side of a baffle remain sonically
connected, but only in that sound waves flow around the outer edge of
that baffle.  Baffles are kept small to prevent interference
with any developing proto-neutron star (PNS) instability.  In current
models, we place the outer radius of the baffle in the range of about
4--8~km, which permits a timestep of about 2--5$ \times 10^{-7}$~s.

\subsection{Neutrino Transport}

Because supernova neutrinos, in general, cannot be described by an
equilibrium distribution function, neutrino transport is the most
difficult component of a supernova model to implement. A solution
requires a complete phase-space description of each neutrino's
position and momentum and, thus, a solution of the six-dimensional
Boltzmann transport equation or some reasonable approximation of it.
This high dimensionality causes the transport portion of a model to
have the largest share of computational cost in execution time,
computer memory, and I/O.

The Boltzmann transport equation can be expressed in terms of the
radiation intensity, $I = I(\epsilon, {\bf x}, {\pmb \Omega}, t),$
where $\epsilon$ is the energy of a neutrino, ${\bf x}$ its position,
and ${\pmb \Omega}$ the solid angle into which the neutrino radiation
is directed. In terms of $I$, the Newtonian form can be expressed as
\begin{equation}\label{eq:bte}
\frac{1}{c} \frac{\partial I}{\partial t} +
{\pmb \Omega} \cdot {\pmb \nabla} I + \sum_i
a_{i} \frac{\partial I}{\partial p_{i}} =
\left( \frac{\partial f}{\partial t} \right)_{{\rm coll.}},
\end{equation}
where $a_{i}$ is the $i^{th}$ component of the matter acceleration,
$p_{i}$ the $i^{th}$ component of the momentum of the neutrino, and
$c$ the speed of light.  The right-hand side of Eq.\ (\ref{eq:bte})
lumps together the contributions from all interactions that a neutrino
might experience and is collectively referred to as the collision
integral.

In V2D, we implement a fully two-dimensional, multi-group flux-limited
diffusion scheme.  This scheme extends our earlier work
\cite{MBHLSV,SSS} and is accomplished by taking the zeroth angular
moment of the Eq.\ (\ref{eq:bte}) to yield the following neutrino
monochromatic energy equation in the co-moving frame
\begin{equation}\label{eq:bte0}
\frac{\partial E_{\epsilon}}{\partial t} +
{\pmb \nabla} \cdot \left( E_{\epsilon} {\bf v} \right) +
{\pmb \nabla} \cdot {\bf F}_{\epsilon} -
\epsilon \frac{\partial}{\partial \epsilon} 
\left( {\mathsf P}_{\epsilon}:
{\pmb \nabla} {\bf v} \right) = {\mathbb S}_{\epsilon},
\end{equation}
where $E_{\epsilon}$ is the zeroth angular moment of $I$ and is also
the neutrino energy density per unit energy interval at position ${\bf
x}$ and time $t$.  The first angular moment of $I$ is given by ${\bf
F}_{\epsilon}$, which is also the neutrino energy flux per unit energy
interval. The quantities ${\mathsf P}_{\epsilon}$ and ${\mathbb
S}_{\epsilon}$ are as previously defined (${\mathsf P}_{\epsilon}$
being additionally the second angular moment of $I$). The expression
${\mathsf P}_{\epsilon}:{\pmb \nabla} {\bf v}$ indicates
contraction in both indices of the second-rank tensors ${\mathsf
P}_{\epsilon}$ and ${\pmb \nabla} {\bf v}$.  A corresponding
equation describes the antineutrinos, which can be expressed by
substituting the antineutrino energy density $\bar{E}_{\epsilon}$, and
corresponding higher moments,  for
each instance of $E_{\epsilon}$ and {\it vice versa} in
Eq.~(\ref{eq:bte0}). This pair of equations is solved for each neutrino
energy $\epsilon$ and neutrino flavor over the entire computational
domain.  We currently track electronic, muonic, and tauonic neutrinos.

The right-hand side of Eq.~(\ref{eq:bte0}) can be expressed as
\begin{eqnarray}\label{eq:source}
{\mathbb S}_{\epsilon} \equiv
S_\epsilon 
\left(1-\frac{\alpha}{\epsilon^3}E_\epsilon \right)
- c \kappa^a_\epsilon E_\epsilon
+\left(1-\frac{\alpha}{{\epsilon}^3}
E_{\epsilon} \right)
\epsilon   \int d\epsilon^\prime
G(\epsilon,\epsilon^\prime)
\left(1-\frac{\alpha}{{\epsilon^\prime}^3}
\bar{E}_{\epsilon^\prime} \right)
\nonumber \\
+ \left(1-\frac{\alpha}{\epsilon^3}E_\epsilon \right)
c \int d\epsilon^\prime 
\kappa^{\rm in}(\epsilon,\epsilon^\prime)
E_{\epsilon^\prime}
- E_{\epsilon}
c \int d\epsilon^\prime 
\kappa^{\rm out}(\epsilon,\epsilon^\prime)
\left(1-\frac{\alpha}{{\epsilon^\prime}^3}
E_{\epsilon^\prime} \right)
\label{eq:src_mono_neut_rad_eng},
\end{eqnarray}
where $S_{\epsilon}$ and $\kappa^a_{\epsilon}$ are the neutrino
emissivities and absorption opacities, respectively, for electron capture
processes---a reaction that is only significant for electron
neutrinos.  Production of neutrino-antineutrino pairs is accounted for
in the term containing the pair production kernel
$G(\epsilon,\epsilon^\prime)$.  Finally, the last pair of terms
accounts for non-conservative scattering, processes that can scatter
neutrinos into an energy state $\epsilon$ from energy state
$\epsilon^\prime$ with opacity $\kappa^{\rm
in}(\epsilon,\epsilon^\prime)$ and the reverse process with opacity
$\kappa^{\rm out}(\epsilon,\epsilon^\prime)$.  Once again, the
analogous expressions for antineutrinos can be expressed with the
substitutions noted above. The phase-space factor $\alpha$ equals
$(hc)^3/4\pi$ for neutrinos, where $h$ is Planck's constant.

Equation~(\ref{eq:bte0}) is closed using Levermore and Pomraning's
prescription for flux-limited diffusion~\cite{LP}, which allows us to
express ${\bf F}_{\epsilon} = -D_{\epsilon} {\pmb \nabla}
E_{\epsilon}$, where $D_{\epsilon}$ is a ``variable'' diffusion
coefficient that yields the correct fluxes for the diffusion and
free-streaming limits and an approximate solution in the intermediate
regime.  This prescription also provides the elements of the
radiation-pressure tensor ${\mathsf P}_{\epsilon}$, which closes the
system.

Since the neutrino CFL condition is far too restrictive to permit an
explicit transport solution, we use a purely implicit method. The
equations comprising each neutrino-antineutrino species are assembled
into linear-system form. Blocking terms arising from Fermi-Dirac statistical
restrictions on final neutrino states---the $1 - \alpha
E_{\epsilon}/\epsilon^3$ terms in Eq.~(\ref{eq:source})---make this a
system of non-linear equations.  Fortunately, the system is sparse,
making it amenable to an iterative solution. A nested
procedure is used, employing Newton-Krylov methods \cite{SSS}.  In the
innermost loop, a linearized system is solved in the pre-conditioned
Krylov-subspace.  The outer loop uses a Newton-Raphson iteration to
resolve the non-linearities.  Besides being an effective general
procedure for sparse systems, our implementation of parallel
pre-conditioners also insures that the solver can run effectively on
large-scale parallel computing architectures---the chief reason our
code exhibits its high degree of scalability.

\subsection{Pre-collapse Progenitor Models and Initial Conditions}

For a progenitor model, we employ the widely used Woosley and Weaver
\cite{ww} S15S7b2 $15 M_{\odot}$ progenitor.  The initial model is
collapsed using RH1D, our 1-D Newtonian Lagrangean radiation
hydrodynamics code. Immediately prior to core bounce, the iron core,
the silicon shell, and a portion of the oxygen shell are zoned into a
256 radial-mass-zone mesh with zoning that is tuned to yield a high
spatial resolution grid in the proto-neutron star and the inner 200~km
of the collapsed core.  This zoning sets up a radial grid that is
compatible with subsequent 2-D Eulerian simulations.  In both the 1-D
collapse and the subsequent 2-D evolution with V2D, the
neutrino-energy spectrum (range, 0--375 MeV) is discretized into
20 energy groups with group widths increasing geometrically with
energy to resolve the Fermi surface of the electrons and neutrinos in
the developing proto-neutron star.  The initial values for $T$,
$\rho$, and $Y_e$ are interpolated from the original S15S7B2 data
onto the Langragean mass grid, and the initial radial coordinates of
each mass shell are computed consistently with density.

\subsection{Nuclear Microphysics and the Equation of State}

Stellar-core collapse simulations require an equation of state (EOS)
that handles a density range of roughly $10^5$--$10^{15}$ g ${\rm
cm^{-3}}$, a temperature range of 0.1--25~MeV, and an
electron-fraction range of 0.0--0.5.  The EOS also must handle
different regimes of equilibrium states.  Throughout most of the core,
the material is in nuclear statistical equilibrium (NSE), though
matter in the silicon shell and beyond does not attain NSE until the
bounce-shock wave passes through it.

For supernova simulations, we use the Lattimer-Swesty EOS
\cite{LS,LLPR} in tabular form.  The thermodynamic quantities are
tabulated in terms of independent variables: density, $\rho$;
temperature, $T$; and electron fraction, $Y_e$.  We have tabulated
this EOS in a thermodynamically consistent way according to the
prescription in Swesty \cite{FDSTCT}. We refer to this combination
collectively as LS-TCT. Although the Lattimer-Swesty EOS is commonly
used, and tabulations of it are also common, most tabular
interpolations are not constructed to {\em guarantee} thermodynamic
consistency.  When a non-thermodynamically-consistent scheme is used,
spurious entropy can be generated or lost. 

\subsection{Neutrino Microphysics}

The energetics of a core-collapse supernova are dominated by neutrinos
and, therefore, it is necessary to have accurate opacities
and rates for the various important neutrino processes.

The collection of reactions that are important, or possibly important,
to the supernova problem is large and evolving. Of undisputed
importance is electron capture by protons and protons bound in nuclei,
neutrino production through electron-positron pair annihilation, and
non-conservative neutrino-electron scattering.  The major contributors
to neutrino opacity are coherent scattering of neutrinos from nuclei
and conservative scattering from free nucleons.

The models presented here contain neutrino microphysics as described
in Bruenn \cite{bruenn85}, with two exceptions.  The neutrino-nucleus
scattering opacity has been modified to take into account the form
factor introduced by Burrows, Mazurek, and Lattimer \cite{bml}, and we
have neglected the effect of neutrino-antineutrino annihilation.
Full neutrino-electron scattering, as presently described, is
included in the code but is not turned on in the models described
here.  Additional effects, including nucleon recoil, ion-ion
correlations, and flavor-changing reactions, are under active
development as additions to these baseline models.

\section{Parallel Implementation}

\begin{figure}[h]
\includegraphics[width=35pc]{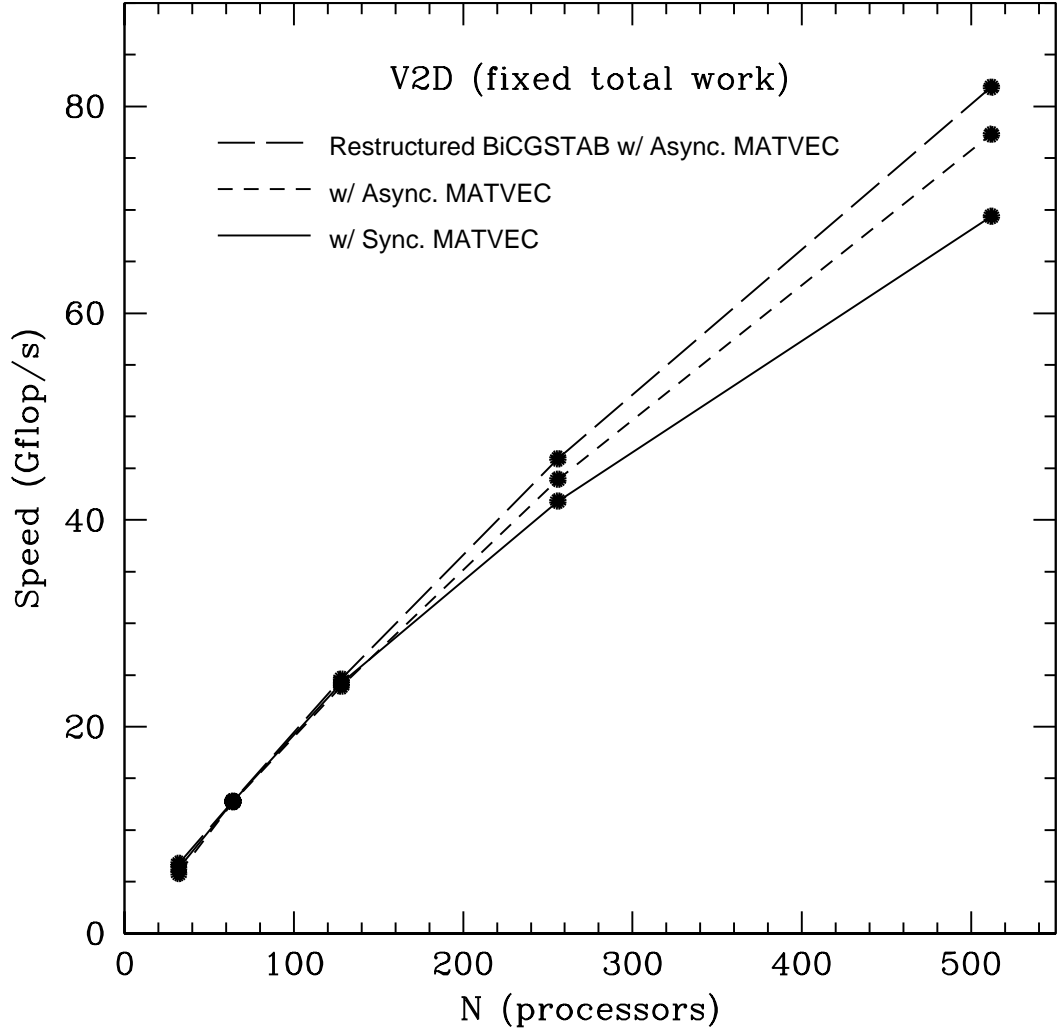}
\caption{\label{fig:f1}Parallel speedup of V2D on the NERSC IBM-SP
(seaborg).}
\end{figure}

The numerical solution of the radiation-hydrodynamics equations
delineated in the prior sections is accomplished via the V2D code
previously mentioned.  The size of the problem we solve
necessitates massively parallel computing resources.  Since we solve a
long-timescale problem, it is necessary that we achieve
strong-scaling, {\em i.e.,} we wish the fixed-size problem to scale
well to large numbers of processors to reduce our wall-clock time to
solution.  Also, the number of variables in the problem requires a
large amount of memory.  Parallelism in V2D is achieved via spatial
domain decomposition of the 2-D spatial domain into a logically
Cartesian topology of 2-D subdomains.  Under this decomposition, our
finite-difference algorithms for the hydrodynamic and
radiation-transport equations require a limited set of communication
patterns.  The evaluation of the finite-difference stencils for both
sets of equations requires only nearest-neighbor message-passing with
this decomposition.  Unfortunately, global reduction operations
are needed in a few instances.  The calculation of the timestep size
$\Delta t$ requires a global reduction to determine the
minimum CFL time for the entire domain.
The iterative solution, by Newton-Krylov methods, of the
implicitly-differenced diffusion equations requires global reductions
to evaluate vector inner products. In the
BiCGSTAB algorithm, which is often used to solve the linear
systems in Newton-Krylov methods, this can mean
multiple global reductions per iteration.  This can impose
a bottleneck to scalability when carrying out simulations on large
numbers of processors.  To reduce this bottleneck, we have
developed a restructured variant of the BiCGSTAB algorithm that is
algebraically equivalent, but requiring only a single global
reduction per iteration, and which also allows for better loop
optimization.  The effect of this improvement can be seen in 
Fig.~\ref{fig:f1}, where we plot the parallel speedup of V2D on a supernova
model on {\em seaborg}, the IBM-SP at NERSC.

The major floating point cost of the Newton-Krylov algorithm is expended
in the multiplication of the Jacobian operator for the
non-linear diffusion equation with a vector.  This operation requires
only nearest-neighbor communication to evaluate the
finite-difference stencil of the divergence operator.
We amortize this cost by performing the nearest neighbor-communication
asynchronously while carrying out the portions of the matrix-vector
multiply operation corresponding to the interior zones of each
subdomain.  This yields a further improvement in scalability as seen in
Fig.~\ref{fig:f1}.

V2D's data input and output is accomplished in parallel using the
parallel HDF5 libraries, which in turn rely on the MPI-I/O interface.
The entire simulation does parallel collective reads and writes into a
single HDF5 file.  This strategy ensures that checkpoint and restart
files are independent of processor count.  I/O is typically costly,
and, therefore, we checkpoint relatively infrequently.  We have also
found that parallel I/O can become an impediment to scaling beyond
1024 processors, and we are investigating strategies to mitigate this
problem.

\section{A Few Scientific Discoveries\ldots}

Using V2D, we have carried out a simulations of the convectively
unstable post-bounce epoch of a core-collapsed 15$M_\odot$ star.  As
mentioned previously, models were evolved through the collapse epoch
using RH1D. Immediately prior to core bounce, when the central density
is about $10^{14}$~g~${\rm cm^{-3}}$, remapping to 2-D spherical
polar coordinates takes place, keeping radial zoning unaltered. For
the S15S7b model and $K=180$ MeV in the LS EOS, this corresponds to a
time of 233.087 ms.  The model is subsequently evolved further with
V2D.  This allows the shock to develop naturally on the Eulerian mesh
and eliminates the possibility of spurious shocks in the quasi-static
post-bounce core that have been observed when the 1D-to-2D transition
is made at a later time.

\begin{figure}[h]
\includegraphics[width=20pc]{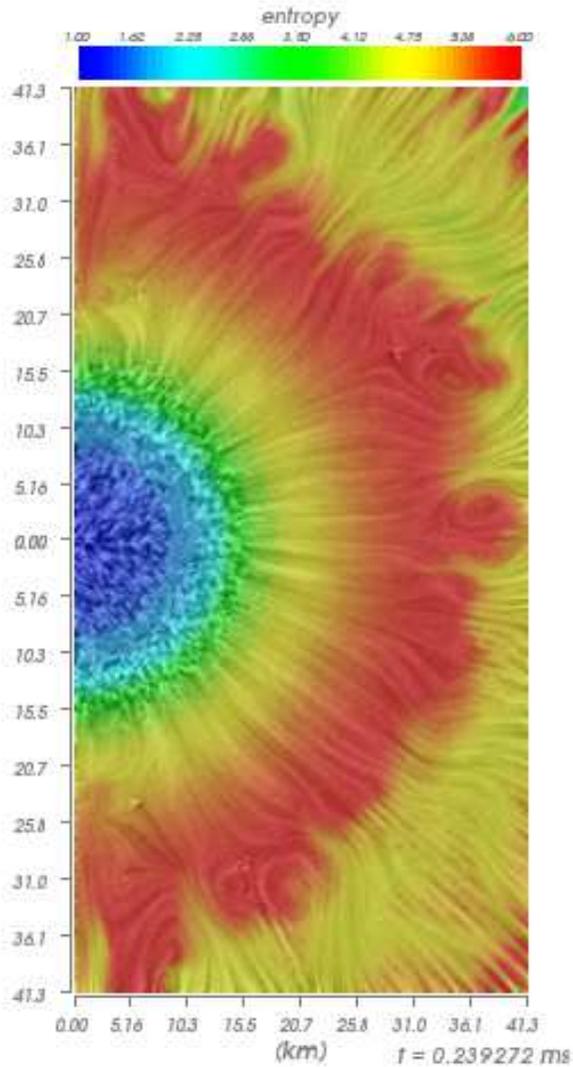}
\caption{\label{fig:f2} The development of Rayleigh-Taylor
initiated convection in a region with negative entropy
gradient.  The color denotes the entropy per baryon while
the streaks are a texture map that show the instantaneous
structure of the velocity field.}
\end{figure}

\begin{figure}[h]
\includegraphics[width=35pc]{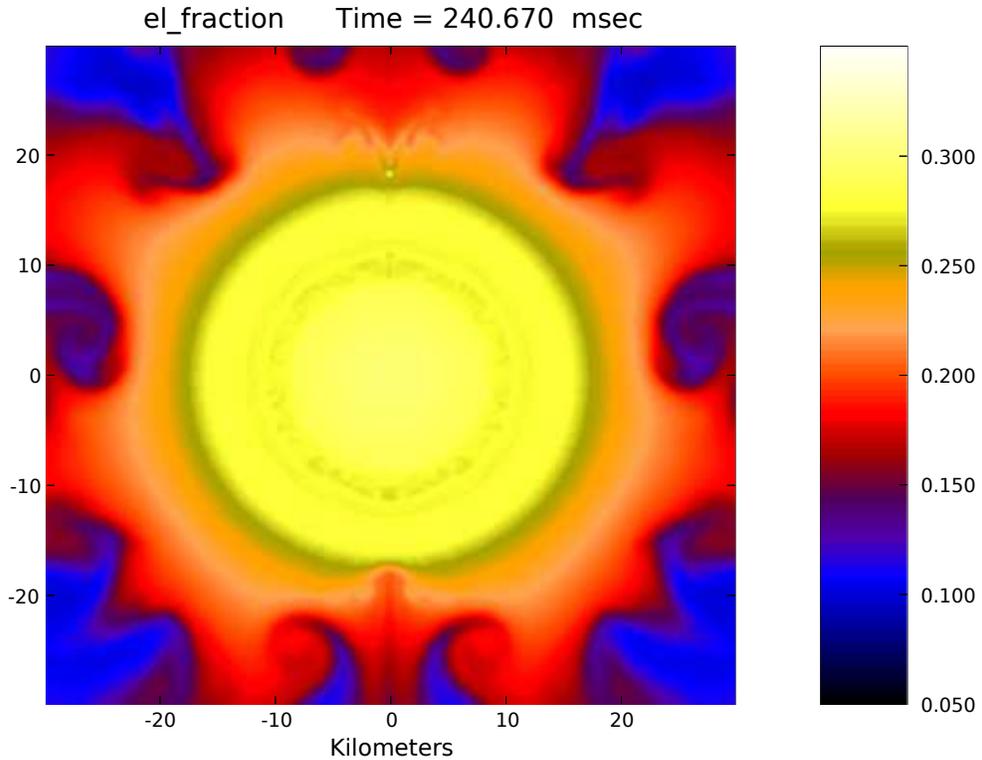}
\caption{\label{fig:f3} The electron fraction in 
the region below the neutrinosphere during the 
brief period of vigorous convection.}
\end{figure}

An interesting discovery we have made is that by making the 
1D-to-2D transition prior to bounce, fluid-instabilities 
develop earlier after the shock breakout stage.  This is 
illustrated in Fig.~\ref{fig:f2}, which shows a 
Rayleigh-Taylor instability
developing in a region with a negative radial entropy gradient
as a result of a weakening shock wave.
This convective instability develops within 5--6 ms of the
beginning of the simulation and core bounce.  This is
substantially earlier than reported by other groups who have their
started 2-D simulations after the prompt shock has stalled. The
instability develops in the region where the material is optically
thick to the bulk of the neutrinos. The convective circulation in this
region causes a brief enhancement to the neutrino luminosity by
advecting upward neutrinos that are ``trapped'' in the fluid on a
diffusion timescale much longer than the fluid dynamic timescale.
When the fluid containing the neutrinos reaches lower densities,
they are able to escape, since their opacity drops
with density.  This convective circulation also drags neutron
rich material downward into the core as seen in Fig.~\ref{fig:f3},
which shows the electron fraction of the material.

This period of vigorous circulation below neutrinosphere lasts only
briefly and does not play a long term role in enhancing the
luminosity.  In fact, after 10 ms, the net flow of neutrinos in this
region is downwards, since they are advected deeper into the core by
high density matter that is quasi-statically settling after passing
through the nearly-stalled accretion shock.  This vigorous convection
in the optically thick regions reduces the entropy gradient, after which
convection in this region diminishes.

\begin{figure}[h]
\includegraphics[width=35pc]{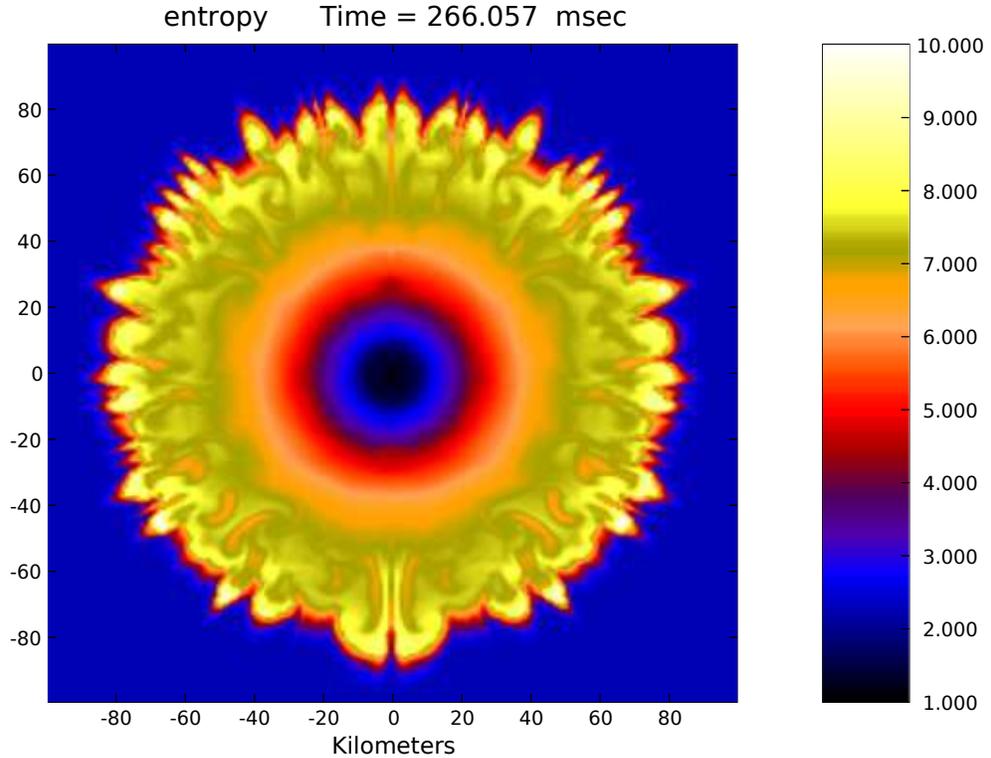}
\caption{\label{fig:f4} The entropy in 
the region below the neutrinosphere during the 
brief period of vigorous convection.}
\end{figure}

In this particular model, the stalled accretion shock does 
not revitalize on the timescale of this simulation (about
33 ms).  Convection persists in the region above the
neutrinosphere but below the stalled shock as is illustrated in 
Fig.~\ref{fig:f4}.   
One can also see that by this time, the convection below the
neutrinosphere has halted after it has stabilized the entropy gradient
out to a radius of approximately 40 km. Whether there is
sufficient reheating taking place in the convective region above
the neutrinosphere to revitalize the shock in the long term, is the
subject of our continuing investigations.

\section{Conclusions and Future Directions}

Through our simulations, we have found that the onset of fluid
instabilities may occur earlier than previously thought.  These
instabilities occur as a result of the negative entropy gradient left
in the wake of a weakening shock wave.  We have also found that, in
this model, 
vigorous protoneutron star convection does not persist for any
significant time after the entropy gradient has been
stabilized.  Our results also indicate that shock revitalization 
by means of convective reheating enhancements does not occur on 
timescales of a few tens of milliseconds.

Our research into this mechanism is continuing at a brisk pace.  We
are moving beyond this baseline model to examine the effects of
various microphysics changes within the model including the variation
of the nuclear force parameters within the equation of state.  Another
area in which we are accelerating our research is in the development
of 3-D models.  It is unclear whether the convective structure we
currently see in 2-D models will persist in 3-D.  In addition, we hope
to explore the effects of rotation with 3-D models.  These studies will
continue to occupy us beyond terascale computing and well into the
petascale era.

\section*{Acknowledgments}
The authors thank numerous people
who have aided our supernova modeling
efforts. Foremost on this list are TSI team
members Jim Lattimer, Ken DeNisco, Amy Irwin,
Chris Tartamella, and Clint Young (SUNY Stony
Brook); Ed Bachta and Polly Baker (Indiana
University at Indianapolis); Dennis Smolarski
(Santa Clara University); John Blondin (North
Carolina State University).  We also thank
members of some of the SciDAC funded ISIC
and SAPP teams that have contributed technologies
that have made our efforts easier: Micah Beck and
Scott Atchley (LOCI Lab, University of Tennessee
Knoxville); Xiaowen Xin and Terence Critchlow
(SDM ISIC Team, LLNL); Dan Reynolds and Carol
Woodward (TOPs ISIC Team, LLNL).  Finally, we 
acknowledge the many members of the NERSC staff
who have worked to help us carry out the
computations described in this paper including
Horst Simon, Francesca Verdier, Wes Bethel, David Skinner,
Richard Gerber, John Shalf, Eli Dart, and Brent
Draney.

We gratefully acknowledge the support of the U.S.\ Department of Energy,
through SciDAC Award DE-FC02-01ER41185, by which this work was funded.
We are also grateful to the National Energy Research Scientific
Computing Center (NERSC) for computational support.

\end{document}